\documentclass[aps,prl,10pt,reprint,superscriptaddress]{revtex4-2}

\usepackage{natbib}
\usepackage{amsmath,amsfonts,amssymb}
\usepackage{xcolor,graphicx}
\usepackage{tikz}
\usepackage{braket}
\usepackage{dcolumn}
\usepackage{bm}
\usepackage{subfig}
\usepackage{float}
\usepackage[pdftex]{hyperref}
\usepackage{csquotes}
\hypersetup{
	colorlinks = true,
	linkcolor = blue,
	anchorcolor = blue,
	citecolor = red,
	filecolor = blue,
	urlcolor = blue}
\begin{document}
\title{Cauchy-Riemann beams}

\author{H. M. Moya-Cessa}
\affiliation{Instituto Nacional de Astrofísica Óptica y Electrónica, Calle Luis Enrique Erro No. 1\\ Santa María Tonantzintla, Puebla, 72840, Mexico}

\author{I. Ramos-Prieto}
\email[e-mail:\,]{iran@inaoep.mx}
\affiliation{Instituto Nacional de Astrofísica Óptica y Electrónica, Calle Luis Enrique Erro No. 1\\ Santa María Tonantzintla, Puebla, 72840, Mexico}

\author{D. Sánchez-de-la-Llave}
\affiliation{Instituto Nacional de Astrofísica Óptica y Electrónica, Calle Luis Enrique Erro No. 1\\ Santa María Tonantzintla, Puebla, 72840, Mexico}

\author{U. Ruíz}
\affiliation{Instituto Nacional de Astrofísica Óptica y Electrónica, Calle Luis Enrique Erro No. 1\\ Santa María Tonantzintla, Puebla, 72840, Mexico}

\author{V. Arrizón}
\affiliation{Instituto Nacional de Astrofísica Óptica y Electrónica, Calle Luis Enrique Erro No. 1\\ Santa María Tonantzintla, Puebla, 72840, Mexico}

\author{F. Soto-Eguibar}
\affiliation{Instituto Nacional de Astrofísica Óptica y Electrónica, Calle Luis Enrique Erro No. 1\\ Santa María Tonantzintla, Puebla, 72840, Mexico}

\date{\today}

\begin{abstract}
	By using operator techniques, we solve the paraxial wave equation for a field given by the multiplication of a Gaussian function and an entire function. The latter possesses a unique property, being an eigenfunction of the {\it perpendicular} Laplacian with a zero eigenvalue, a consequence of the Cauchy-Riemann equations. We demonstrate, both theoretically and experimentally, the inherent rotation of this field during its propagation. The explanation for these rotations lies in the utilization of the quantum (Bohm) potential. The simplicity of this outcome reveals promising prospects: it enables the analytical deduction of the Fraunhofer or Fresnel diffraction pattern. In essence, this means that obtaining the Fresnel or Fourier transform of a function satisfying the Cauchy-Riemann equations becomes a straightforward task.
\end{abstract}
\maketitle

\textbf{Introduction.} Over the past five decades, the scientific community has seen the emergence of numerous solutions to the paraxial equation, which are crucial in the study of wave propagation. These solutions have explored various coordinate systems, ranging from specific field forms to more general ones. While examples of these solutions can be found in a wide range of references~\cite{Durnin_1987,Gori_1987,Durham,Gutierrez_2000,HollowGauss,Bandres_2004,Bandres_2004a,Gutierrez_2005,Bandres_2007,Kotlyar_2007,KiselevReview,Karimi_2007,Cartesian,Bandres_2008,Elliptical}, they by no means constitute an exhaustive account of the field. One particularly distinctive approach, outlined in a prior study~\cite{Abramochkin_1993}, involves breaking down the paraxial equation into a system of two equations. This approach treats the field's magnitude and phase as separate entities and proposes a structurally stable ansatz for the field intensity. This method results in a solution encompassing functions where the complex amplitude can be represented by a complex entire function multiplied by a Gaussian distribution.

In this research, we introduce a novel approach aimed at deriving closed-form solutions to the paraxial equation. Our approach builds upon the well-established fact that entire functions are solutions to the Laplace equation. The solutions we present are equivalent to those presented by Abramochkin et al.~\cite{Abramochkin_1993}, where complex amplitudes of fields were initially represented by entire functions multiplied by a real Gaussian function at $z = 0$. Leveraging concepts from quantum optics operators, we formulate closed-form expressions that precisely describe the propagation of these fields. One intriguing characteristic of these beams is their innate ability to undergo self-transformation during a Fourier transform operation, provided an axis rotation is allowed. To shed light on this phenomenon, we propose a fresh interpretation of the field's rotation based on the quantum Bohm potential. Our innovative approach not only contributes to a deeper understanding of these wave fields, but also extends their practical applications. By providing a new perspective on their behavior and properties, our research aims to enhance comprehension and open up new avenues for utilizing these wave phenomena across various fields.

It is well known that an analytic function $f(x+iy)$ satisfies the Cauchy-Riemann equations and, in turn, is an eigenfunction of the transverse Laplacian, i.e., $\nabla_{\perp}^2f(x+iy) = 0$, with $\nabla^2_\perp = \frac{\partial^2}{\partial_x^2} + \frac{\partial^2}{\partial^2_y}$ denoting the transverse component of the Laplacian \cite{Churchill,Schleich_2023}. Undoubtedly, when the function $f$ exhibits multivalued behavior within specific regions or at certain points in the complex plane, it remains eligible for classification as an analytic function, subject to specific limitations. In light of these considerations, and in the context of optical scalar fields, we have chosen to designate such fields as Cauchy-Riemann beams~(CRB). These beams denote functions that are differentiable within a certain region, potentially spanning the entire complex plane, and satisfying the Cauchy-Riemann equations.  It is worth noting that Abramochkin et al. (among others, e.g., Kotlyar et al.~\cite{Kotlyar_2023}) had already observed indications of these fields, initially labeling them as Spiral-type beams~\cite{Abramochkin_1993,Abramochkin_1996}.

Let $f(x+iy)$ be an eigenfunction of the transverse Laplacian with zero eigenvalue; thus, this function is a solution to the paraxial equation
\begin{equation}\label{0020}
	\nabla_\perp^2 E\left( x,y,z\right) +2ik\frac{\partial E\left( x,y,z\right) }{\partial z}=0,
\end{equation}
where $k = \frac{2\pi}{\lambda}$ is the wave number.\\
Certainly, similar to other adiffractional beams like Bessel beams~\cite{Durnin_1987}, these solutions are not physically realizable because they are not square-integrable. In light of this, drawing inspiration from Bessel-Gauss beams~\cite{Gori_1987}, we seek a generalized version that incorporates a Gaussian factor to render the field square-integrable. It is essential to emphasize that not all analytic functions, when multiplied by such Gaussian factors, become square-integrable. To achieve this goal, we write the paraxial equation as a Schrödinger-like equation, $\frac{\partial E\left(x,y,z\right) }{\partial z}=\frac{i}{2k}\nabla_\perp^2 E\left(x,y,z\right)$, whose formal solution can be writte as: $E\left(x,y,z\right)=\exp\left(i \frac{z}{2k}  \nabla_\perp^2\right) E(x,y,0)$,
being $E(x,y,0)$ the initial field at $z=0$. In the subsequent analysis, we adopt an approach involving operators less commonly employed in physical optics~\cite{Stoler_1980}. This approach, however, proves advantageous in obtaining solutions to the paraxial equation when an initial condition is provided. As a result, in Cartesian coordinates, the field at location $z$ is calculated as:
\begin{equation}\label{0050}
	E\left(x,y,z\right)=\exp\left[-\frac{i}{2k} z \left(\hat{p}_x^2+\hat{p}_y^2 \right) \right] E(x,y,0),
\end{equation}
where we have introduced the operators $\hat{p}_x=-i\frac{\partial}{\partial x}$ and $\hat{p}_y=-i\frac{\partial}{\partial y}$,  which obey the following commutation relations: $[x,\hat{p}_x]= [y,\hat{p}_y]= i$ and $[x,y]=[x,\hat{p}_y]= [y,\hat{p}_x]=[\hat{p}_x,\hat{p}_y]=0$. Now, we write the initial condition as: $E(x,y,0)=\exp\left[-g\left(x^2+y^2\right)\right]f(x+iy)$, with $g$ taking values in the complex number domain ($g\in\mathbb{C}$). By substituting this initial condition into Eq.~\eqref{0050}, we obtain $E\left(x,y,z\right)=\exp\left[-\frac{i}{2k} z \left(\hat{p}_x^2+\hat{p}_y^2 \right) \right]\exp\left[-g\left(x^2+y^2\right)\right]f(x+iy)$. As subsequent step, we introduce the identity operator $\hat{I}$, expressed as $\hat{I}=e^{\frac{i}{2k} z \left(\hat{p}_x^2+\hat{p}_y^2 \right)}e^{-\frac{i}{2k} z \left(\hat{p}_x^2+\hat{p}_y^2 \right)}$. Therefore, the equation that describes the field $E\left(x,y,z\right)$, incorporating this identity operator, is given as follows
\begin{equation}\label{007}
	\begin{split}
		E\left(x,y,z\right)&=\exp\left[-\frac{i}{2k} z \left(\hat{p}_x^2+\hat{p}_y^2 \right) \right]\exp\left[-g\left(x^2+y^2\right)\right]\\
		&\times\exp\left[\frac{i}{2k} z \left(\hat{p}_x^2+\hat{p}_y^2 \right) \right]\exp\left[-\frac{i}{2k} z \left(\hat{p}_x^2+\hat{p}_y^2 \right) \right]\\&\times f(x+iy).
	\end{split}
\end{equation}
The previous equation has two fundamental ingredients. The first one is that the set of operators $\hat{p}_q^2$, $q^2$, and $q\hat{p}_q+\hat{p}_qq$ (with $q = x, y$) is closed under commutation. Consequently, by using the Hadamard Lemma~\cite{RossmannW,Hall_2013}, it is possible to demonstrate that, $
	e^{-i\frac{Z}{2}\hat{p}_q^2}e^{-gq^2}e^{i\frac{Z}{2}\hat{p}_q^2}
	=e^{-g[q^2-Z(q\hat{p}_q+\hat{p}_qq)+Z^2\hat{p}_q^2]}$, with  $q=x,y$,
where we re-scaled the propagation distance to $Z=z/k$. The second, and more significant aspect (which serves as the inspiration for the work's title), is that an analytic function $f(x+iy)$ satisfy the Cauchy-Riemann equations, and acts as an eigenfunction of the transverse Laplacian operator, with zero eigenvalue. Therefore, from Eq.~\eqref{007}, it follows that
\begin{equation}\label{E9}
	\begin{split}
		E\left(x,y,z\right)&=e^{-g[x^2-Z(x\hat{p}_x+\hat{p}_xx)+Z^2\hat{p}_x^2]}\\
		&\times e^{-g[y^2-Z(y\hat{p}_y+\hat{p}_yy)+Z^2\hat{p}_y^2]} f(x+iy).
	\end{split}
\end{equation}
The commutation-closed characteristic of the operator set permits the factorization of the exponential operator in the aforementioned equation~\cite{WN_1964},
\begin{equation}\label{factorizacion}
	e^{-g(q^2-Z[q\hat{p}_q+\hat{p}_qq]+Z^2\hat{p}_q^2)}=e^{\alpha(z)q^2}e^{\beta(z)(q\hat{p}_q+\hat{p}_qq)}e^{\gamma(z)\hat{p}_q^2},
\end{equation}
where $\alpha(z)=\frac{-g}{w(z)}$, $\beta(z)=-\frac{\pi}{4}-\frac{i}{2}\ln[iw(z)]$, and  $\gamma(z)=-\frac{gZ^2}{w(z)}$, with $w(z) = 2igZ+1$, for $q = x,y$ respectively. Therefore, as a direct consequence of $e^{\gamma(z)(\hat{p}_x^2+\hat{p}_y^2)}f(x+iy) = f(x+iy)$, by substituting Eq.~\eqref{factorizacion} into Eq.~\eqref{E9}, we can represent the solution as: $E\left(x,y,z\right)=e^{\alpha(z)\left(x^2+y^2\right)}e^{\beta(z)\left(x\hat{p}_x+\hat{p}_xx+y\hat{p}_y+\hat{p}_yy\right)}f({x+iy})$.
Finally, the last exponential in the above equation is the well-known squeeze operator that may be applied to the analytic function to give
\begin{equation}\label{0140}
	E\left(x,y,z\right)=\frac{\exp\left[-\frac{g\left(x^2+y^2\right)}{w(z)}\right]}{w(z)}f\left(\frac{x+iy}{w(z)}\right).
\end{equation}
The preceding result provides a comprehensive understanding of the propagation of fields that satisfy the Cauchy-Riemann equations, which are modulated at $z=0$ by either a Gaussian function, a quadratic-phase function, or both. This representation proves to be versatile and generic, rendering it suitable for a wide array of scenarios in which these equations are applicable.
\begin{figure}[H]
	\centering
	\includegraphics[width = .85\linewidth]{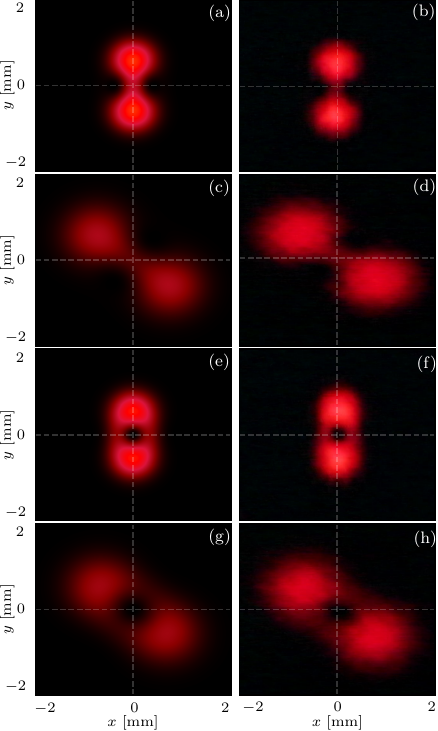}
	\caption{Intensity distribution of the field for (a)-(b)-(e)-(f) at $z=0$, and (c)-(d)-(g)-(h) at $z=.5\,\text{m}$ , respectively. The left column represents a numerical simulation based on Eq.~\eqref{0140}, while the right column corresponds to an experimental realization. Experimental parameters: $g = 1.25 \times 10^7\,\text{m}^{-2}$, $T = 0.0008\,\text{m}$, and $\lambda= 633\,\text{nm}$, all within a viewing window in millimeters.}
	\label{Fig_1}
\end{figure}
As illustrative examples, we consider the entire functions $f_1(x+iy)=\cos\left(\frac{2\pi}{T}(x+iy)\right)$ and $f_2(x+iy)=J_1 \left(\frac{2\pi}{T}(x+iy)\right)$, where $J_1\left(\zeta\right)$ is the Bessel function with $n=1$. Fig.~\ref{Fig_1} shows both, the numerical and the experimental intensity distribution of the fields given by Eq.~\eqref{0140}, on the planes $z=0.0 \, \text{m}$ and $z=0.5 \, \text{m}$, with the parameters $T=0.0008 \, \text{m}$, the $x$ and $y$ coordinates ranging from $-0.2 \, \text{cm}$ to $0.2 \, \text{cm}$, and the constant $g=1.25 \times 10^7 \, \text{m}^{-2}$. For the experimental fields, we use a collimated He-Ne laser ($\lambda=633 \, \text{nm}$) and a phase-only spatial light modulator (PLUTO, Holoeye GmbH), addressed by means of a synthetic phase hologram in a 4f-optical system~\cite{Arrizon07}. The left and right column of the Fig.~\ref{Fig_1} shows the numerical and experimental normalized intensity of the generated fields $E_1(x,y,z)$ and $E_2(x,y,z)$, respectively. As can be noted the Figs. 1(c) and 1(g), which are the intensity of the fields on $z=0.5 \, \text{m}$, present a rotation respect to the fields in Figs. 1(a) and 1(e), which are the intensity of the fields on $z=0.0 \, \text{m}$, respectively.

\textbf{Fresnel diffraction.} To examine in greater detail the field distribution and its characteristics when the parameter $g$ takes complex values, i.e., $g = g_R + ig_I$ (with $g_R > 0, \ g_I \geq 0$), the quadratic phase term associated to $g_I$ can be regarded as a thin lens with a focal length of $1/2g_I$. In this sense, the field at $Z = 1/2g_I$ is determined according to Eq.~\eqref{0140} as follows:
\begin{equation}\label{0151}
	\begin{split}
		E \left(x,y,\frac{1}{2g_I} \right)&= \frac{-ig_I}{g_R} \exp \left[-\frac{g_I^2}{g_R} \left(x^2+y^2 \right) \right]\\&
		\times \exp \left[ig_I \left(x^2+y^2 \right)  \right] f\left[\frac{-ig_I}{g_R}(x+iy)\right].
	\end{split}
\end{equation}
Comparing the above equation with the field predicted by the Fresnel integral of diffraction, one can deduce that the field given by Eq.~\eqref{0140} at the plane $Z=0$, with $g_R$ taking a real positive value, is a self-transforming field under the Fourier transform operator. In general, the field at $Z=1/(2g_I)$ is a scaled version of the field at $Z=0$ (without the thin lens factor) and undergoes an axis rotation of $- \pi/2$ rad. An interesting example arises when $g_I=g_R$. In this case, the field at $Z=1/(2g_I)$ is a replica of the field at $Z=0$ (without the quadratic phase term) with an axis rotation of $- \pi/2 $ rad. Next, we analyze the field properties in the far field region.

	{\bf Fraunhofer diffraction.} Continuing along the same line of thought, when the condition $2Z \lvert g_R+ig_R \rvert \gg 1$ is met, indicating far field diffraction, and replacing $Z=\lambda z/2 \pi$, subsequent to some algebraic manipulations, Eq.~\eqref{0140} can be reformulated as follows:
\begin{equation}\label{0170}
	\begin{split}
		E\left(x,y,z \right)&=\frac{-i \pi}{\lambda z g^*}
		\exp\left[\frac{i \pi\left(x^2+y^2 \right)}{\lambda z}-\frac{\pi^2}{g^*}\frac{(x^2+y^2)}{\lambda^2z^2}\right]\\
		&\times f\left[\frac{\pi}{\lambda z} \left( \frac{-i x}{g^*} + \frac{y}{g} \right) \right],
	\end{split}
\end{equation}
where the symbol $^*$ denotes the complex conjugate. Therefore, comparing the above equation with the field given by the Fraunhofer integral of diffraction, one can conclude that the Fourier transform of the field described by Eq.~\eqref{0140} at the plane $Z=0$ is given by
\begin{equation}
	\begin{split}
		\mathcal{F}\left\{e^{-g\left(x^2+y^2\right)}f(x+iy)\right\}&=\frac{\pi}{g^*} \exp\left[\frac{-\pi^2}{g^*} \left(v_{x}^{2}+v_{y}^{2}\right)\right]\\&\times f\left[\pi \left( \frac{-i v_x}{g^*} + \frac{v_y}{g} \right) \right],
	\end{split}
\end{equation}
where $v_{x}$ and $v_{y}$ are the spatial frequencies in Cartesian coordinates. Aside from the axis rotation of the entire function, it is remarkable that this family of functions is self-transforming under the Fourier transform operator. To the best of our knowledge, this is also a novel result.\\
Finally, we note that the total axis rotation the field undergoes as it propagates across the whole $z$-axis is $-\pi$ rad. However, the axis rotation as the field propagates from $z=0$ up to the far field region is $-\pi /2$ rad, for $g_I=0$, and tends to $-\pi$ rad when $g_I$ is much larger than $g_R$. For the case of negative $g_I$, the field rotates less than $-\pi/2$ rad as it propagates from $z=0$ to the far field zone. In general, the rate of rotation of the axis is not linear with $z$, being zero in the far field.

\begin{figure}[H]
	\includegraphics[width = .85\linewidth]{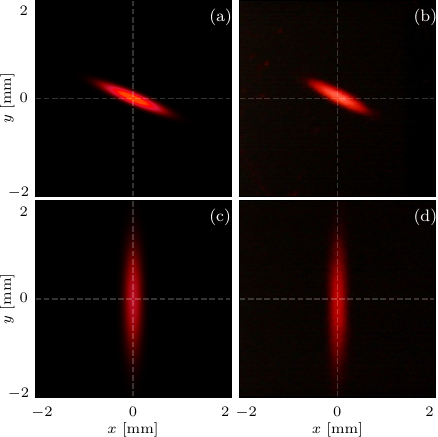}
	\caption{The left column represents the intensity of the field for $f(x+iy)=\exp\left[\eta(x+iy)^2\right]$, while the right column shows the experimental results. The parameters used in the experimental setup are as follows: $T = 8 \times 10^{-4}\,\text{m}$, $g = \frac{12}{T^{2}}(1+i)$, and $\eta = \frac{2}{3}g$. Panels (a)-(b) depict the intensity distribution at $z = 0.0\,\text{m}$, and panels (c)-(d) display the intensity at $z = 0.45\,\text{m}$ using Eq.~\eqref{0140}.}
	\label{Fig_3}
\end{figure}

\begin{figure}[H]
	\includegraphics[width = .85\linewidth]{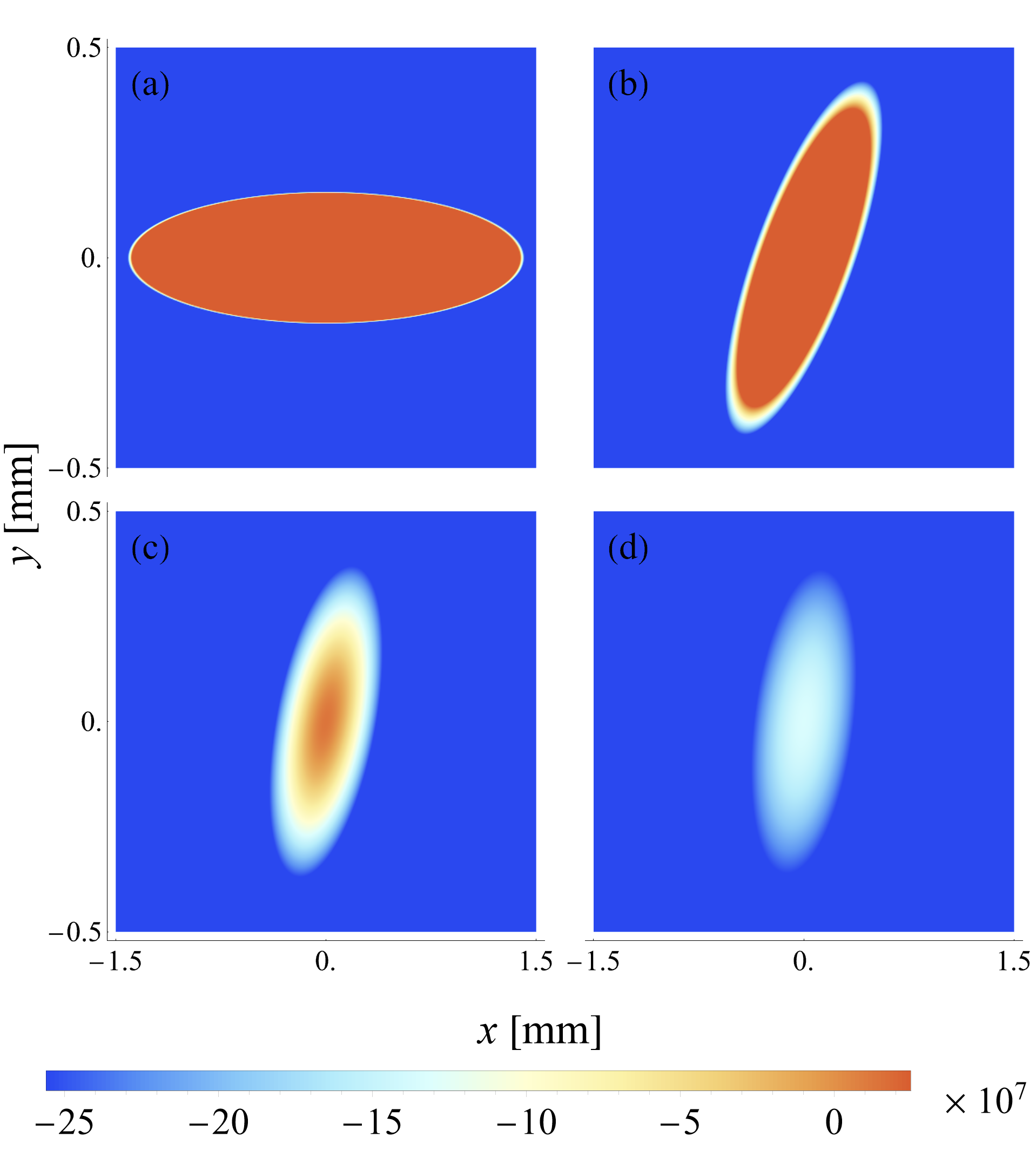}
	\caption{The Bohm potential for the Cauchy-Riemann field given in \eqref{0140a} at distances from $0.0$ to $3.0\times10^{-7}$ in increments of $10^{-7}$. The values of the parameters are $g=1.25\times 10^7$ and $\eta=1.0\times 10^7$.}
	\label{potBohm}
\end{figure}

\textbf{Bohm formalism.} It is well known that Airy waves bend while they propagate \cite{Siviloglou_2007,Rozenman_2019,Hojman_2021,Silva_2023}. Furthermore, their Bohm trajectories \cite{Bohm_1952} may be demonstrated in hydrodynamic systems \cite{Rozenman_2023}, where the quantum potential has been shown to be linear. The rotation suffered by the Cauchy-Riemann beams during propagation, may also be attributed to the so-called quantum potential. In the Bohm formalism, we have that by writing
\begin{equation}\label{AmpPhase}
	E(x,y,z)=A(x,y,z)\exp[iS(x,y,z)],
\end{equation}
the differential equations  that obey the amplitude and the phase  are
$\frac{1}{2}S_x^2+\frac{1}{2}S_y^2+V_B+S_t=0$,
where as we are dealing with free space propagation have omitted the standard potential, and
\begin{equation}\label{Prob}
	\frac{\partial A}{\partial z}+
	S_x\frac{\partial A}{\partial x}
	+S_y \frac{\partial A}{\partial y}
	+\frac{1}{2} \left(S_{xx} A + S_{yy} A\right)=0,
\end{equation}
where the sub-indices represent partial derivatives, and  the dependence on $x,\;y,\;z$ has been omitted. The quantum or Bohm potential is
\begin{equation}\label{bohmpot}
	V_B(x,y,z)=-\frac{1}{2 } \frac{\nabla_{\perp}^2 A(x,y,z)}{A(x,y,z)}.
\end{equation}
The Bohm potential for the Cauchy-Riemann beams, Eq.~\eqref{0140}, is given by
\begin{equation}
	\begin{split}
		Q(x,y,z)&=\frac{2 g}{w(z)}-\frac{2 g^2 \left(x^2+y^2\right)}{w(z)^2}
		+\frac{2 g (x+i y)}{w(z)^2}\\&\times
		\frac{f'\left(\frac{x+i y}{w(z)}\right)}{f\left(\frac{x+i y}{w(z)}\right)},
	\end{split}
\end{equation}
where the prime denotes derivative with respect to the argument.\\
For the functions used in Fig. \ref{Fig_1}, namely cosine and Bessel functions, the quantum potential exists and produces the effect of rotating the field, but its expressions are complex. A function of the form $f(x+iy)=\exp\left[\eta(x+iy)^2\right]$,
besides delivering a simple quantum potential, that produces the same rotating effect as seen in Fig.~\ref{Fig_3} theoretically and experimentally, is a good example to show that not any entire function has square integrable properties, as for $|\eta|\ge|g|$ the field has infinite energy. The propagated field in this case is given by
\begin{equation}\label{0140a}
	E\left(x,y,z\right)=\frac{e^{-g\frac{x^2+y^2}{R(z)}e^{-i\Phi_{G}}}}{R(z)}e^{\eta\frac{(x+iy)^2}{R^2(z)}e^{-2i\Phi_{G}}}e^{-i\Phi_{G}},
\end{equation}
where $\Phi_{G}=\arctan(2gZ)$ is the so-called Gouy phase, that may explained with the help of the quantum Bohm potential \cite{Moya_2022} and $R(z)=|w(z)|$. From the above equation, we get
\begin{equation}
	\begin{split}
		&A(x,y,z)=\frac{\exp\left[-\frac{g(x^2+y^2)\cos\Phi_G}{R(z)}\right]}{R}\\&
		\times\exp\left[\frac{\eta[(x^2-y^2)\cos(2\Phi_G)+2xy\sin(2\Phi_G)]}{R^2(z)}\right],
	\end{split}
\end{equation}
and
\begin{align}
	 & S(x,y,z)=-\Phi_G+\frac{g(x^2+y^2)\sin\Phi_G}{R(z)}
	\nonumber                                             \\ &
	+\frac{\eta[2xy\cos(2\Phi_G)-(x^2-y^2)\sin(2\Phi_G)]}{R^2(z)},
\end{align}
from where we readily may find the quantum potential
\begin{align}\label{BP}
	 & V_B=-2(x^2+y^2)\left(\frac{\eta^2}{R^4}+\frac{g^2}{R^2}\cos^2\Phi_G\right)+\frac{2g}{R}\cos\Phi_G
	\nonumber                                                                                            \\  &
	+\frac{4g\eta}{R^3}\cos\Phi_G[\cos(2\Phi_G)(x^2-y^2)+2xy\sin(2\Phi_G)],
\end{align}
where we have omitted the dependence on $z$ of the function $R(z)$. The Bohm potential, responsible for the rotation of the beams, is  plotted in Fig.~\ref{potBohm} for several propagation distances. From (\ref{BP}) it may be seen that the Bohm potential represents a GRIN medium \cite{ASENJO2021126947} that changes with the propagation distance.

\textbf{Conclusions.} We have demonstrated a solution to the paraxial equation using an unconventional approach in the field of paraxial optics, specifically by employing the operator technique from quantum mechanics, and more precisely, quantum optics. This approach is grounded in the fact that $\nabla_{\perp}^2f(x+iy)=0$.  Consequently, it follows that $f(x+iy)$ serves as an eigenfunction of the Laplacian operator in two dimensions with an  eigenvalue zero, an attribute stemming from $f(x+iy)$ satisfying the Cauchy-Riemann equations. In addition, we have explored specific cases of the general result provided by Eq.~\eqref{0140}, which holds significant interest in the field of diffractive optics. It allows for the straightforward and direct determination of Fresnel and Fraunhofer diffraction for a wide range of initial conditions when considering $g$ in the complex number domain. It is worth noting that experimental evidence supports the observation that this solution exhibits rotation. Furthermore, we have developed the Bohm quantum potential to elucidate the reason behind the observed rotation in these solutions.

%

\begin{thebibliography}{33}%
	\makeatletter
	\providecommand \@ifxundefined [1]{%
	 \@ifx{#1\undefined}
	}%
	\providecommand \@ifnum [1]{%
	 \ifnum #1\expandafter \@firstoftwo
	 \else \expandafter \@secondoftwo
	 \fi
	}%
	\providecommand \@ifx [1]{%
	 \ifx #1\expandafter \@firstoftwo
	 \else \expandafter \@secondoftwo
	 \fi
	}%
	\providecommand \natexlab [1]{#1}%
	\providecommand \enquote  [1]{``#1''}%
	\providecommand \bibnamefont  [1]{#1}%
	\providecommand \bibfnamefont [1]{#1}%
	\providecommand \citenamefont [1]{#1}%
	\providecommand \href@noop [0]{\@secondoftwo}%
	\providecommand \href [0]{\begingroup \@sanitize@url \@href}%
	\providecommand \@href[1]{\@@startlink{#1}\@@href}%
	\providecommand \@@href[1]{\endgroup#1\@@endlink}%
	\providecommand \@sanitize@url [0]{\catcode `\\12\catcode `\$12\catcode
	  `\&12\catcode `\#12\catcode `\^12\catcode `\_12\catcode `\%12\relax}%
	\providecommand \@@startlink[1]{}%
	\providecommand \@@endlink[0]{}%
	\providecommand \url  [0]{\begingroup\@sanitize@url \@url }%
	\providecommand \@url [1]{\endgroup\@href {#1}{\urlprefix }}%
	\providecommand \urlprefix  [0]{URL }%
	\providecommand \Eprint [0]{\href }%
	\providecommand \doibase [0]{https://doi.org/}%
	\providecommand \selectlanguage [0]{\@gobble}%
	\providecommand \bibinfo  [0]{\@secondoftwo}%
	\providecommand \bibfield  [0]{\@secondoftwo}%
	\providecommand \translation [1]{[#1]}%
	\providecommand \BibitemOpen [0]{}%
	\providecommand \bibitemStop [0]{}%
	\providecommand \bibitemNoStop [0]{.\EOS\space}%
	\providecommand \EOS [0]{\spacefactor3000\relax}%
	\providecommand \BibitemShut  [1]{\csname bibitem#1\endcsname}%
	\let\auto@bib@innerbib\@empty
	\bibitem [{\citenamefont {Durnin}(1987)}]{Durnin_1987}%
	  \BibitemOpen
	  \bibfield  {author} {\bibinfo {author} {\bibfnamefont {J.}~\bibnamefont
	  {Durnin}},\ }\bibfield  {title} {\bibinfo {title} {Exact solutions for
	  nondiffracting beams. {I}. the scalar theory},\ }\href
	  {https://doi.org/10.1364/JOSAA.4.000651} {\bibfield  {journal} {\bibinfo
	  {journal} {J. Opt. Soc. Am. A}\ }\textbf {\bibinfo {volume} {4}},\ \bibinfo
	  {pages} {651} (\bibinfo {year} {1987})}\BibitemShut {NoStop}%
	\bibitem [{\citenamefont {Gori}\ \emph {et~al.}(1987)\citenamefont {Gori},
	  \citenamefont {Guattari},\ and\ \citenamefont {Padovani}}]{Gori_1987}%
	  \BibitemOpen
	  \bibfield  {author} {\bibinfo {author} {\bibfnamefont {F.}~\bibnamefont
	  {Gori}}, \bibinfo {author} {\bibfnamefont {G.}~\bibnamefont {Guattari}},\
	  and\ \bibinfo {author} {\bibfnamefont {C.}~\bibnamefont {Padovani}},\
	  }\bibfield  {title} {\bibinfo {title} {Bessel-{G}auss beams},\ }\href
	  {https://doi.org/https://doi.org/10.1016/0030-4018(87)90276-8} {\bibfield
	  {journal} {\bibinfo  {journal} {Optics Communications}\ }\textbf {\bibinfo
	  {volume} {64}},\ \bibinfo {pages} {491} (\bibinfo {year} {1987})}\BibitemShut
	  {NoStop}%
	\bibitem [{\citenamefont {Caron}\ and\ \citenamefont
	  {Potvliege}(1999)}]{Durham}%
	  \BibitemOpen
	  \bibfield  {author} {\bibinfo {author} {\bibfnamefont {C.}~\bibnamefont
	  {Caron}}\ and\ \bibinfo {author} {\bibfnamefont {R.}~\bibnamefont
	  {Potvliege}},\ }\bibfield  {title} {\bibinfo {title} {Bessel-modulated
	  {G}aussian beams with quadratic radial dependence},\ }\href
	  {https://doi.org/https://doi.org/10.1016/S0030-4018(99)00174-1} {\bibfield
	  {journal} {\bibinfo  {journal} {Optics Communications}\ }\textbf {\bibinfo
	  {volume} {164}},\ \bibinfo {pages} {83} (\bibinfo {year} {1999})}\BibitemShut
	  {NoStop}%
	\bibitem [{\citenamefont {Guti\'{e}rrez-Vega}\ \emph
	  {et~al.}(2000)\citenamefont {Guti\'{e}rrez-Vega}, \citenamefont
	  {Iturbe-Castillo},\ and\ \citenamefont {Ch\'{a}vez-Cerda}}]{Gutierrez_2000}%
	  \BibitemOpen
	  \bibfield  {author} {\bibinfo {author} {\bibfnamefont {J.~C.}\ \bibnamefont
	  {Guti\'{e}rrez-Vega}}, \bibinfo {author} {\bibfnamefont {M.~D.}\ \bibnamefont
	  {Iturbe-Castillo}},\ and\ \bibinfo {author} {\bibfnamefont {S.}~\bibnamefont
	  {Ch\'{a}vez-Cerda}},\ }\bibfield  {title} {\bibinfo {title} {Alternative
	  formulation for invariant optical fields: {M}athieu beams},\ }\href
	  {https://doi.org/10.1364/OL.25.001493} {\bibfield  {journal} {\bibinfo
	  {journal} {Opt. Lett.}\ }\textbf {\bibinfo {volume} {25}},\ \bibinfo {pages}
	  {1493} (\bibinfo {year} {2000})}\BibitemShut {NoStop}%
	\bibitem [{\citenamefont {Cai}\ \emph {et~al.}(2003)\citenamefont {Cai},
	  \citenamefont {Lu},\ and\ \citenamefont {Lin}}]{HollowGauss}%
	  \BibitemOpen
	  \bibfield  {author} {\bibinfo {author} {\bibfnamefont {Y.}~\bibnamefont
	  {Cai}}, \bibinfo {author} {\bibfnamefont {X.}~\bibnamefont {Lu}},\ and\
	  \bibinfo {author} {\bibfnamefont {Q.}~\bibnamefont {Lin}},\ }\bibfield
	  {title} {\bibinfo {title} {Hollow {G}aussian beams and their propagation
	  properties},\ }\href {https://doi.org/10.1364/OL.28.001084} {\bibfield
	  {journal} {\bibinfo  {journal} {Opt. Lett.}\ }\textbf {\bibinfo {volume}
	  {28}},\ \bibinfo {pages} {1084} (\bibinfo {year} {2003})}\BibitemShut
	  {NoStop}%
	\bibitem [{\citenamefont {Bandres}\ and\ \citenamefont
	  {Guti\'{e}rrez-Vega}(2004)}]{Bandres_2004}%
	  \BibitemOpen
	  \bibfield  {author} {\bibinfo {author} {\bibfnamefont {M.~A.}\ \bibnamefont
	  {Bandres}}\ and\ \bibinfo {author} {\bibfnamefont {J.~C.}\ \bibnamefont
	  {Guti\'{e}rrez-Vega}},\ }\bibfield  {title} {\bibinfo {title}
	  {Ince--{G}aussian beams},\ }\href {https://doi.org/10.1364/OL.29.000144}
	  {\bibfield  {journal} {\bibinfo  {journal} {Opt. Lett.}\ }\textbf {\bibinfo
	  {volume} {29}},\ \bibinfo {pages} {144} (\bibinfo {year} {2004})}\BibitemShut
	  {NoStop}%
	\bibitem [{\citenamefont {Bandres}\ \emph {et~al.}(2004)\citenamefont
	  {Bandres}, \citenamefont {Guti\'{e}rrez-Vega},\ and\ \citenamefont
	  {Ch\'{a}vez-Cerda}}]{Bandres_2004a}%
	  \BibitemOpen
	  \bibfield  {author} {\bibinfo {author} {\bibfnamefont {M.~A.}\ \bibnamefont
	  {Bandres}}, \bibinfo {author} {\bibfnamefont {J.~C.}\ \bibnamefont
	  {Guti\'{e}rrez-Vega}},\ and\ \bibinfo {author} {\bibfnamefont
	  {S.}~\bibnamefont {Ch\'{a}vez-Cerda}},\ }\bibfield  {title} {\bibinfo {title}
	  {Parabolic nondiffracting optical wave fields},\ }\href
	  {https://doi.org/10.1364/OL.29.000044} {\bibfield  {journal} {\bibinfo
	  {journal} {Opt. Lett.}\ }\textbf {\bibinfo {volume} {29}},\ \bibinfo {pages}
	  {44} (\bibinfo {year} {2004})}\BibitemShut {NoStop}%
	\bibitem [{\citenamefont {Guti\'{e}rrez-Vega}\ and\ \citenamefont
	  {Bandres}(2005)}]{Gutierrez_2005}%
	  \BibitemOpen
	  \bibfield  {author} {\bibinfo {author} {\bibfnamefont {J.~C.}\ \bibnamefont
	  {Guti\'{e}rrez-Vega}}\ and\ \bibinfo {author} {\bibfnamefont {M.~A.}\
	  \bibnamefont {Bandres}},\ }\bibfield  {title} {\bibinfo {title}
	  {Helmholtz--{G}auss waves},\ }\href {https://doi.org/10.1364/JOSAA.22.000289}
	  {\bibfield  {journal} {\bibinfo  {journal} {J. Opt. Soc. Am. A}\ }\textbf
	  {\bibinfo {volume} {22}},\ \bibinfo {pages} {289} (\bibinfo {year}
	  {2005})}\BibitemShut {NoStop}%
	\bibitem [{\citenamefont {Bandres}\ and\ \citenamefont
	  {Guti\'{e}rrez-Vega}(2007{\natexlab{a}})}]{Bandres_2007}%
	  \BibitemOpen
	  \bibfield  {author} {\bibinfo {author} {\bibfnamefont {M.~A.}\ \bibnamefont
	  {Bandres}}\ and\ \bibinfo {author} {\bibfnamefont {J.~C.}\ \bibnamefont
	  {Guti\'{e}rrez-Vega}},\ }\bibfield  {title} {\bibinfo {title} {Airy-{G}auss
	  beams and their transformation by paraxial optical systems},\ }\href
	  {https://doi.org/10.1364/OE.15.016719} {\bibfield  {journal} {\bibinfo
	  {journal} {Opt. Express}\ }\textbf {\bibinfo {volume} {15}},\ \bibinfo
	  {pages} {16719} (\bibinfo {year} {2007}{\natexlab{a}})}\BibitemShut {NoStop}%
	\bibitem [{\citenamefont {Kotlyar}\ \emph {et~al.}(2007)\citenamefont
	  {Kotlyar}, \citenamefont {Skidanov}, \citenamefont {Khonina},\ and\
	  \citenamefont {Soifer}}]{Kotlyar_2007}%
	  \BibitemOpen
	  \bibfield  {author} {\bibinfo {author} {\bibfnamefont {V.~V.}\ \bibnamefont
	  {Kotlyar}}, \bibinfo {author} {\bibfnamefont {R.~V.}\ \bibnamefont
	  {Skidanov}}, \bibinfo {author} {\bibfnamefont {S.~N.}\ \bibnamefont
	  {Khonina}},\ and\ \bibinfo {author} {\bibfnamefont {V.~A.}\ \bibnamefont
	  {Soifer}},\ }\bibfield  {title} {\bibinfo {title} {Hypergeometric modes},\
	  }\href {https://doi.org/10.1364/OL.32.000742} {\bibfield  {journal} {\bibinfo
	   {journal} {Opt. Lett.}\ }\textbf {\bibinfo {volume} {32}},\ \bibinfo {pages}
	  {742} (\bibinfo {year} {2007})}\BibitemShut {NoStop}%
	\bibitem [{\citenamefont {Kiselev}(2007)}]{KiselevReview}%
	  \BibitemOpen
	  \bibfield  {author} {\bibinfo {author} {\bibfnamefont {A.~P.}\ \bibnamefont
	  {Kiselev}},\ }\bibfield  {title} {\bibinfo {title} {Localized light waves:
	  Paraxial and exact solutions of the wave equation (a review)},\ }\href
	  {https://doi.org/10.1134/S0030400X07040200} {\bibfield  {journal} {\bibinfo
	  {journal} {Optics and Spectroscopy}\ }\textbf {\bibinfo {volume} {102}},\
	  \bibinfo {pages} {603} (\bibinfo {year} {2007})}\BibitemShut {NoStop}%
	\bibitem [{\citenamefont {Karimi}\ \emph {et~al.}(2007)\citenamefont {Karimi},
	  \citenamefont {Zito}, \citenamefont {Piccirillo}, \citenamefont {Marrucci},\
	  and\ \citenamefont {Santamato}}]{Karimi_2007}%
	  \BibitemOpen
	  \bibfield  {author} {\bibinfo {author} {\bibfnamefont {E.}~\bibnamefont
	  {Karimi}}, \bibinfo {author} {\bibfnamefont {G.}~\bibnamefont {Zito}},
	  \bibinfo {author} {\bibfnamefont {B.}~\bibnamefont {Piccirillo}}, \bibinfo
	  {author} {\bibfnamefont {L.}~\bibnamefont {Marrucci}},\ and\ \bibinfo
	  {author} {\bibfnamefont {E.}~\bibnamefont {Santamato}},\ }\bibfield  {title}
	  {\bibinfo {title} {Hypergeometric-{G}aussian modes},\ }\href
	  {https://doi.org/10.1364/OL.32.003053} {\bibfield  {journal} {\bibinfo
	  {journal} {Opt. Lett.}\ }\textbf {\bibinfo {volume} {32}},\ \bibinfo {pages}
	  {3053} (\bibinfo {year} {2007})}\BibitemShut {NoStop}%
	\bibitem [{\citenamefont {Bandres}\ and\ \citenamefont
	  {Guti\'{e}rrez-Vega}(2007{\natexlab{b}})}]{Cartesian}%
	  \BibitemOpen
	  \bibfield  {author} {\bibinfo {author} {\bibfnamefont {M.~A.}\ \bibnamefont
	  {Bandres}}\ and\ \bibinfo {author} {\bibfnamefont {J.~C.}\ \bibnamefont
	  {Guti\'{e}rrez-Vega}},\ }\bibfield  {title} {\bibinfo {title} {Cartesian
	  beams},\ }\href {https://doi.org/10.1364/OL.32.003459} {\bibfield  {journal}
	  {\bibinfo  {journal} {Opt. Lett.}\ }\textbf {\bibinfo {volume} {32}},\
	  \bibinfo {pages} {3459} (\bibinfo {year} {2007}{\natexlab{b}})}\BibitemShut
	  {NoStop}%
	\bibitem [{\citenamefont {Bandres}\ and\ \citenamefont
	  {Guti\'{e}rrez-Vega}(2008{\natexlab{a}})}]{Bandres_2008}%
	  \BibitemOpen
	  \bibfield  {author} {\bibinfo {author} {\bibfnamefont {M.~A.}\ \bibnamefont
	  {Bandres}}\ and\ \bibinfo {author} {\bibfnamefont {J.~C.}\ \bibnamefont
	  {Guti\'{e}rrez-Vega}},\ }\bibfield  {title} {\bibinfo {title} {Circular
	  beams},\ }\href {https://doi.org/10.1364/OL.33.000177} {\bibfield  {journal}
	  {\bibinfo  {journal} {Opt. Lett.}\ }\textbf {\bibinfo {volume} {33}},\
	  \bibinfo {pages} {177} (\bibinfo {year} {2008}{\natexlab{a}})}\BibitemShut
	  {NoStop}%
	\bibitem [{\citenamefont {Bandres}\ and\ \citenamefont
	  {Guti\'{e}rrez-Vega}(2008{\natexlab{b}})}]{Elliptical}%
	  \BibitemOpen
	  \bibfield  {author} {\bibinfo {author} {\bibfnamefont {M.~A.}\ \bibnamefont
	  {Bandres}}\ and\ \bibinfo {author} {\bibfnamefont {J.~C.}\ \bibnamefont
	  {Guti\'{e}rrez-Vega}},\ }\bibfield  {title} {\bibinfo {title} {Elliptical
	  beams},\ }\href {https://doi.org/10.1364/OE.16.021087} {\bibfield  {journal}
	  {\bibinfo  {journal} {Opt. Express}\ }\textbf {\bibinfo {volume} {16}},\
	  \bibinfo {pages} {21087} (\bibinfo {year} {2008}{\natexlab{b}})}\BibitemShut
	  {NoStop}%
	\bibitem [{\citenamefont {Abramochkin}\ and\ \citenamefont
	  {Volostnikov}(1993)}]{Abramochkin_1993}%
	  \BibitemOpen
	  \bibfield  {author} {\bibinfo {author} {\bibfnamefont {E.}~\bibnamefont
	  {Abramochkin}}\ and\ \bibinfo {author} {\bibfnamefont {V.}~\bibnamefont
	  {Volostnikov}},\ }\bibfield  {title} {\bibinfo {title} {Spiral-type beams},\
	  }\href {https://doi.org/https://doi.org/10.1016/0030-4018(93)90406-U}
	  {\bibfield  {journal} {\bibinfo  {journal} {Optics Communications}\ }\textbf
	  {\bibinfo {volume} {102}},\ \bibinfo {pages} {336} (\bibinfo {year}
	  {1993})}\BibitemShut {NoStop}%
	\bibitem [{\citenamefont {Brown}\ and\ \citenamefont
	  {Churchill}(2009)}]{Churchill}%
	  \BibitemOpen
	  \bibfield  {author} {\bibinfo {author} {\bibfnamefont {J.~W.}\ \bibnamefont
	  {Brown}}\ and\ \bibinfo {author} {\bibfnamefont {R.~V.}\ \bibnamefont
	  {Churchill}},\ }\href@noop {} {\emph {\bibinfo {title} {Complex Variables and
	  Applications}}},\ \bibinfo {edition} {eighth}\ ed.\ (\bibinfo  {publisher}
	  {McGraw-Hill Higher Education},\ \bibinfo {address} {Boston, MA},\ \bibinfo
	  {year} {2009})\BibitemShut {NoStop}%
	\bibitem [{\citenamefont {Schleich}\ \emph {et~al.}(2023)\citenamefont
	  {Schleich}, \citenamefont {Tk{\'a}{\v{c}}ov{\'a}},\ and\ \citenamefont
	  {Happ}}]{Schleich_2023}%
	  \BibitemOpen
	  \bibfield  {author} {\bibinfo {author} {\bibfnamefont {W.~P.}\ \bibnamefont
	  {Schleich}}, \bibinfo {author} {\bibfnamefont {I.}~\bibnamefont
	  {Tk{\'a}{\v{c}}ov{\'a}}},\ and\ \bibinfo {author} {\bibfnamefont
	  {L.}~\bibnamefont {Happ}},\ }\bibinfo {title} {Insights into complex
	  functions},\ in\ \href {https://doi.org/10.1007/978-3-031-32469-7_5} {\emph
	  {\bibinfo {booktitle} {Sketches of Physics: The Celebration Collection}}},\
	  \bibinfo {editor} {edited by\ \bibinfo {editor} {\bibfnamefont
	  {R.}~\bibnamefont {Citro}}, \bibinfo {editor} {\bibfnamefont
	  {M.}~\bibnamefont {Lewenstein}}, \bibinfo {editor} {\bibfnamefont
	  {A.}~\bibnamefont {Rubio}}, \bibinfo {editor} {\bibfnamefont {W.~P.}\
	  \bibnamefont {Schleich}}, \bibinfo {editor} {\bibfnamefont {J.~D.}\
	  \bibnamefont {Wells}},\ and\ \bibinfo {editor} {\bibfnamefont {G.~P.}\
	  \bibnamefont {Zank}}}\ (\bibinfo  {publisher} {Springer International
	  Publishing},\ \bibinfo {address} {Cham},\ \bibinfo {year} {2023})\ pp.\
	  \bibinfo {pages} {127--159}\BibitemShut {NoStop}%
	\bibitem [{\citenamefont {Kotlyar}\ \emph {et~al.}(2023)\citenamefont
	  {Kotlyar}, \citenamefont {Kovalev},\ and\ \citenamefont
	  {Abramochkin}}]{Kotlyar_2023}%
	  \BibitemOpen
	  \bibfield  {author} {\bibinfo {author} {\bibfnamefont {V.~V.}\ \bibnamefont
	  {Kotlyar}}, \bibinfo {author} {\bibfnamefont {A.~A.}\ \bibnamefont
	  {Kovalev}},\ and\ \bibinfo {author} {\bibfnamefont {E.~G.}\ \bibnamefont
	  {Abramochkin}},\ }\bibfield  {title} {\bibinfo {title} {Topological charge of
	  propagation-invariant laser beams},\ }\bibfield  {journal} {\bibinfo
	  {journal} {Photonics}\ }\textbf {\bibinfo {volume} {10}},\ \href
	  {https://doi.org/10.3390/photonics10080915} {10.3390/photonics10080915}
	  (\bibinfo {year} {2023})\BibitemShut {NoStop}%
	\bibitem [{\citenamefont {Abramochkin}\ and\ \citenamefont
	  {Volostnikov}(1996)}]{Abramochkin_1996}%
	  \BibitemOpen
	  \bibfield  {author} {\bibinfo {author} {\bibfnamefont {E.}~\bibnamefont
	  {Abramochkin}}\ and\ \bibinfo {author} {\bibfnamefont {V.}~\bibnamefont
	  {Volostnikov}},\ }\bibfield  {title} {\bibinfo {title} {Spiral-type beams:
	  optical and quantum aspects},\ }\href
	  {https://doi.org/https://doi.org/10.1016/0030-4018(95)00640-0} {\bibfield
	  {journal} {\bibinfo  {journal} {Optics Communications}\ }\textbf {\bibinfo
	  {volume} {125}},\ \bibinfo {pages} {302} (\bibinfo {year}
	  {1996})}\BibitemShut {NoStop}%
	\bibitem [{\citenamefont {Stoler}(1981)}]{Stoler_1980}%
	  \BibitemOpen
	  \bibfield  {author} {\bibinfo {author} {\bibfnamefont {D.}~\bibnamefont
	  {Stoler}},\ }\bibfield  {title} {\bibinfo {title} {Operator methods in
	  physical optics},\ }\href {https://doi.org/10.1364/JOSA.71.000334} {\bibfield
	   {journal} {\bibinfo  {journal} {J. Opt. Soc. Am.}\ }\textbf {\bibinfo
	  {volume} {71}},\ \bibinfo {pages} {334} (\bibinfo {year} {1981})}\BibitemShut
	  {NoStop}%
	\bibitem [{\citenamefont {Rossmann}(2002)}]{RossmannW}%
	  \BibitemOpen
	  \bibfield  {author} {\bibinfo {author} {\bibfnamefont {W.}~\bibnamefont
	  {Rossmann}},\ }\href
	  {https://books.google.com/books/about/Lie_Groups.html?id=EjDazZvcquwC} {\emph
	  {\bibinfo {title} {Lie Groups: An Introduction Through Linear Groups}}}\
	  (\bibinfo  {publisher} {Oxford University Press},\ \bibinfo {year}
	  {2002})\BibitemShut {NoStop}%
	\bibitem [{\citenamefont {Hall}(2013)}]{Hall_2013}%
	  \BibitemOpen
	  \bibfield  {author} {\bibinfo {author} {\bibfnamefont {B.~C.}\ \bibnamefont
	  {Hall}},\ }\href {https://doi.org/10.1007/978-1-4614-7116-5_16} {\emph
	  {\bibinfo {title} {Lie Groups, Lie Algebras, and Representations}}}\
	  (\bibinfo  {publisher} {Springer New York},\ \bibinfo {year}
	  {2013})\BibitemShut {NoStop}%
	\bibitem [{\citenamefont {Wei}\ and\ \citenamefont {Norman}(1964)}]{WN_1964}%
	  \BibitemOpen
	  \bibfield  {author} {\bibinfo {author} {\bibfnamefont {J.}~\bibnamefont
	  {Wei}}\ and\ \bibinfo {author} {\bibfnamefont {E.}~\bibnamefont {Norman}},\
	  }\bibfield  {title} {\bibinfo {title} {On global representations of the
	  solutions of linear differential equations as a product of exponentials},\
	  }\href {http://www.jstor.org/stable/2034065} {\bibfield  {journal} {\bibinfo
	  {journal} {Proceedings of the American Mathematical Society}\ }\textbf
	  {\bibinfo {volume} {15}},\ \bibinfo {pages} {327} (\bibinfo {year}
	  {1964})}\BibitemShut {NoStop}%
	\bibitem [{\citenamefont {Arriz\'{o}n}\ \emph {et~al.}(2007)\citenamefont
	  {Arriz\'{o}n}, \citenamefont {Ruiz}, \citenamefont {Carrada},\ and\
	  \citenamefont {Gonz\'{a}lez}}]{Arrizon07}%
	  \BibitemOpen
	  \bibfield  {author} {\bibinfo {author} {\bibfnamefont {V.}~\bibnamefont
	  {Arriz\'{o}n}}, \bibinfo {author} {\bibfnamefont {U.}~\bibnamefont {Ruiz}},
	  \bibinfo {author} {\bibfnamefont {R.}~\bibnamefont {Carrada}},\ and\ \bibinfo
	  {author} {\bibfnamefont {L.~A.}\ \bibnamefont {Gonz\'{a}lez}},\ }\bibfield
	  {title} {\bibinfo {title} {Pixelated phase computer holograms for the
	  accurate encoding of scalar complex fields},\ }\href
	  {https://doi.org/10.1364/JOSAA.24.003500} {\bibfield  {journal} {\bibinfo
	  {journal} {J. Opt. Soc. Am. A}\ }\textbf {\bibinfo {volume} {24}},\ \bibinfo
	  {pages} {3500} (\bibinfo {year} {2007})}\BibitemShut {NoStop}%
	\bibitem [{\citenamefont {Siviloglou}\ \emph {et~al.}(2007)\citenamefont
	  {Siviloglou}, \citenamefont {Broky}, \citenamefont {Dogariu},\ and\
	  \citenamefont {Christodoulides}}]{Siviloglou_2007}%
	  \BibitemOpen
	  \bibfield  {author} {\bibinfo {author} {\bibfnamefont {G.~A.}\ \bibnamefont
	  {Siviloglou}}, \bibinfo {author} {\bibfnamefont {J.}~\bibnamefont {Broky}},
	  \bibinfo {author} {\bibfnamefont {A.}~\bibnamefont {Dogariu}},\ and\ \bibinfo
	  {author} {\bibfnamefont {D.~N.}\ \bibnamefont {Christodoulides}},\ }\bibfield
	   {title} {\bibinfo {title} {Observation of accelerating {A}iry beams},\
	  }\href {https://doi.org/10.1103/PhysRevLett.99.213901} {\bibfield  {journal}
	  {\bibinfo  {journal} {Phys. Rev. Lett.}\ }\textbf {\bibinfo {volume} {99}},\
	  \bibinfo {pages} {213901} (\bibinfo {year} {2007})}\BibitemShut {NoStop}%
	\bibitem [{\citenamefont {Rozenman}\ \emph {et~al.}(2019)\citenamefont
	  {Rozenman}, \citenamefont {Zimmermann}, \citenamefont {Efremov},
	  \citenamefont {Schleich}, \citenamefont {Shemer},\ and\ \citenamefont
	  {Arie}}]{Rozenman_2019}%
	  \BibitemOpen
	  \bibfield  {author} {\bibinfo {author} {\bibfnamefont {G.~G.}\ \bibnamefont
	  {Rozenman}}, \bibinfo {author} {\bibfnamefont {M.}~\bibnamefont
	  {Zimmermann}}, \bibinfo {author} {\bibfnamefont {M.~A.}\ \bibnamefont
	  {Efremov}}, \bibinfo {author} {\bibfnamefont {W.~P.}\ \bibnamefont
	  {Schleich}}, \bibinfo {author} {\bibfnamefont {L.}~\bibnamefont {Shemer}},\
	  and\ \bibinfo {author} {\bibfnamefont {A.}~\bibnamefont {Arie}},\ }\bibfield
	  {title} {\bibinfo {title} {Amplitude and phase of wave packets in a linear
	  potential},\ }\href {https://doi.org/10.1103/PhysRevLett.122.124302}
	  {\bibfield  {journal} {\bibinfo  {journal} {Phys. Rev. Lett.}\ }\textbf
	  {\bibinfo {volume} {122}},\ \bibinfo {pages} {124302} (\bibinfo {year}
	  {2019})}\BibitemShut {NoStop}%
	\bibitem [{\citenamefont {Hojman}\ \emph {et~al.}(2021)\citenamefont {Hojman},
	  \citenamefont {Asenjo}, \citenamefont {Moya-Cessa},\ and\ \citenamefont
	  {Soto-Eguibar}}]{Hojman_2021}%
	  \BibitemOpen
	  \bibfield  {author} {\bibinfo {author} {\bibfnamefont {S.~A.}\ \bibnamefont
	  {Hojman}}, \bibinfo {author} {\bibfnamefont {F.~A.}\ \bibnamefont {Asenjo}},
	  \bibinfo {author} {\bibfnamefont {H.~M.}\ \bibnamefont {Moya-Cessa}},\ and\
	  \bibinfo {author} {\bibfnamefont {F.}~\bibnamefont {Soto-Eguibar}},\
	  }\bibfield  {title} {\bibinfo {title} {Bohm potential is real and its effects
	  are measurable},\ }\href
	  {https://doi.org/https://doi.org/10.1016/j.ijleo.2021.166341} {\bibfield
	  {journal} {\bibinfo  {journal} {Optik}\ }\textbf {\bibinfo {volume} {232}},\
	  \bibinfo {pages} {166341} (\bibinfo {year} {2021})}\BibitemShut {NoStop}%
	\bibitem [{\citenamefont {Silva-Ortigoza}\ and\ \citenamefont
	  {Ortiz-Flores}(2023)}]{Silva_2023}%
	  \BibitemOpen
	  \bibfield  {author} {\bibinfo {author} {\bibfnamefont {G.}~\bibnamefont
	  {Silva-Ortigoza}}\ and\ \bibinfo {author} {\bibfnamefont {J.}~\bibnamefont
	  {Ortiz-Flores}},\ }\bibfield  {title} {\bibinfo {title} {Properties of the
	  {A}iry beam by means of the quantum potential approach},\ }\href
	  {https://doi.org/10.1088/1402-4896/ace2fd} {\bibfield  {journal} {\bibinfo
	  {journal} {Physica Scripta}\ }\textbf {\bibinfo {volume} {98}},\ \bibinfo
	  {pages} {085106} (\bibinfo {year} {2023})}\BibitemShut {NoStop}%
	\bibitem [{\citenamefont {Bohm}(1952)}]{Bohm_1952}%
	  \BibitemOpen
	  \bibfield  {author} {\bibinfo {author} {\bibfnamefont {D.}~\bibnamefont
	  {Bohm}},\ }\bibfield  {title} {\bibinfo {title} {A suggested interpretation
	  of the quantum theory in terms of "hidden" variables. i},\ }\href
	  {https://doi.org/10.1103/PhysRev.85.166} {\bibfield  {journal} {\bibinfo
	  {journal} {Phys. Rev.}\ }\textbf {\bibinfo {volume} {85}},\ \bibinfo {pages}
	  {166} (\bibinfo {year} {1952})}\BibitemShut {NoStop}%
	\bibitem [{\citenamefont {Rozenman}\ \emph {et~al.}(2023)\citenamefont
	  {Rozenman}, \citenamefont {Bondar}, \citenamefont {Schleich}, \citenamefont
	  {Shemer},\ and\ \citenamefont {Arie}}]{Rozenman_2023}%
	  \BibitemOpen
	  \bibfield  {author} {\bibinfo {author} {\bibfnamefont {G.~G.}\ \bibnamefont
	  {Rozenman}}, \bibinfo {author} {\bibfnamefont {D.~I.}\ \bibnamefont
	  {Bondar}}, \bibinfo {author} {\bibfnamefont {W.~P.}\ \bibnamefont
	  {Schleich}}, \bibinfo {author} {\bibfnamefont {L.}~\bibnamefont {Shemer}},\
	  and\ \bibinfo {author} {\bibfnamefont {A.}~\bibnamefont {Arie}},\ }\bibfield
	  {title} {\bibinfo {title} {Observation of {B}ohm trajectories and quantum
	  potentials of classical waves},\ }\href
	  {https://doi.org/10.1088/1402-4896/acb408} {\bibfield  {journal} {\bibinfo
	  {journal} {Physica Scripta}\ }\textbf {\bibinfo {volume} {98}},\ \bibinfo
	  {pages} {044004} (\bibinfo {year} {2023})}\BibitemShut {NoStop}%
	\bibitem [{\citenamefont {Moya-Cessa}\ \emph {et~al.}(2022)\citenamefont
	  {Moya-Cessa}, \citenamefont {Hojman}, \citenamefont {Asenjo},\ and\
	  \citenamefont {Soto-Eguibar}}]{Moya_2022}%
	  \BibitemOpen
	  \bibfield  {author} {\bibinfo {author} {\bibfnamefont {H.~M.}\ \bibnamefont
	  {Moya-Cessa}}, \bibinfo {author} {\bibfnamefont {S.~A.}\ \bibnamefont
	  {Hojman}}, \bibinfo {author} {\bibfnamefont {F.~A.}\ \bibnamefont {Asenjo}},\
	  and\ \bibinfo {author} {\bibfnamefont {F.}~\bibnamefont {Soto-Eguibar}},\
	  }\bibfield  {title} {\bibinfo {title} {Bohm approach to the {G}ouy phase
	  shift},\ }\href {https://doi.org/https://doi.org/10.1016/j.ijleo.2021.168468}
	  {\bibfield  {journal} {\bibinfo  {journal} {Optik}\ }\textbf {\bibinfo
	  {volume} {252}},\ \bibinfo {pages} {168468} (\bibinfo {year}
	  {2022})}\BibitemShut {NoStop}%
	\bibitem [{\citenamefont {Asenjo}\ \emph {et~al.}(2021)\citenamefont {Asenjo},
	  \citenamefont {Hojman}, \citenamefont {Moya-Cessa},\ and\ \citenamefont
	  {Soto-Eguibar}}]{ASENJO2021126947}%
	  \BibitemOpen
	  \bibfield  {author} {\bibinfo {author} {\bibfnamefont {F.~A.}\ \bibnamefont
	  {Asenjo}}, \bibinfo {author} {\bibfnamefont {S.~A.}\ \bibnamefont {Hojman}},
	  \bibinfo {author} {\bibfnamefont {H.~M.}\ \bibnamefont {Moya-Cessa}},\ and\
	  \bibinfo {author} {\bibfnamefont {F.}~\bibnamefont {Soto-Eguibar}},\
	  }\bibfield  {title} {\bibinfo {title} {Propagation of light in linear and
	  quadratic grin media: The bohm potential},\ }\href
	  {https://doi.org/https://doi.org/10.1016/j.optcom.2021.126947} {\bibfield
	  {journal} {\bibinfo  {journal} {Optics Communications}\ }\textbf {\bibinfo
	  {volume} {490}},\ \bibinfo {pages} {126947} (\bibinfo {year}
	  {2021})}\BibitemShut {NoStop}%
	\end{thebibliography}
\end{document}